# A Net Energy-based Analysis for a Climate-constrained Sustainable Energy Transition

Sgouris Sgouridis[*], Ugo Bardi[+], Denes Csala[-]

[*] Institute Center for Smart and Sustainable Systems. Masdar Institute. PO Box 54224. Abu Dhabi, United Arab Emirates. Tel: +971 2 810 9156, Fax: +971 2 810 9901. Email: ssgouridis@masdar.ac.ae and ssgouridis@alum.mit.edu

[+] University of Florence, Department of Earth Sciences, C/o Dipartimento di Chimica, Polo Scientifico di Sesto Fiorentino, 50019, Florence, Italy

Email: ugo.bardi@unifi.it

[-] Institute Center for Smart and Sustainable Systems. Masdar Institute. PO Box 54224. Abu Dhabi, United Arab Emirates. Email: dcsala@masdar.ac.ae

**Summary**

The transition from a fossil-based energy economy to one based on renewable energy is driven by the double challenge of climate change and resource depletion. Building a renewable energy infrastructure requires an upfront energy investment that subtracts from the net energy available to society. This investment is determined by the need to transition to renewable energy fast enough to stave off the worst consequences of climate change and, at the same time, maintain a sufficient net energy flow to sustain the world's economy and population. We show that a feasible transition pathway requires that the rate of investment in renewable energy should accelerate approximately by an order of magnitude if we are to stay within the range of IPCC recommendations.

**Main Text**

The focus on climate change mitigation is on limiting emissions from the combustion of fossil hydrocarbons, overlooking to an extent the provision of sufficient energy resources for a prosperous humanity. Perhaps we should invert this perspective and focus on providing economically sufficient renewable energy (RE) which would imply that less enforcement may be necessary to maintain a large amount of fossil carbon unburned [1].

Historical energy transitions have been slow, typically spanning several decades [2]. However, a sustainable energy transition (SET) from fossil sources to RE is subjected to the constraint that fossil fuels should be phased out fast enough to avoid the worst consequences of climate change. This should be accomplished while providing sufficient net energy to sustain the global economy and accounting for the lower quality of the depleting fossil fuel stocks [3]. In this context a key characteristic of SET becomes critical: it *requires energy* to construct the necessary RE infrastructure and to massively integrate these mostly variable sources in the energy system. At present, the world's energy remains primarily from fossil sources and, as a consequence, *we need energy from fossil fuels to move away from fossil fuels*. We may see this requirement as analogous to "the sower's strategy" [4], the long-established farming practice to save a fraction of the current year's harvest as seeds for the next. Fossil fuels produce no "seed" of their own but we can "sow" what these fuels provide: energy and minerals to create the capital needed for the transition [5]. Yet, withdrawing the "seed" energy reduces net available energy for society - a crucial metric for sustaining our socioeconomic metabolism [6,7]. The challenge is to balance these contradicting needs and complete the SET before fossil fuel depletion makes it impossible or climate change damages become crippling.

Such an energy metabolism perspective simplifies a notable confusion in the discussion of RE potentials [8] as it can provide a range for the RE investment effort (the "seed") and objectively inform policy formation using back-casting to match the required net energy implied by population levels. In order to systematically assess the transition progress, we propose four quantifiable guidelines building on ideas from economics [9, 10]:

*I. The impacts of energy production during SET do not exceed the long-run ecosystem carrying and assimilation capacity.*

*II. Per capita net available energy remains above a desired level that satisfies societal needs at any point during SET and without disruptive discontinuities in its rate of change.*

*III. The investment rate for the installation of renewable generation and consumption capital stock is sufficient to create a sustainable energy supply basis before the non-renewable safely recoverable resources are exhausted.*

*IV. Financial commitments of future consumption (debt) should be limited by future energy availability.*

Guideline I limits the emissions from fossil fuels on the basis of the carrying capacity of the ecosystem, mainly in regard to greenhouse gas (GHG) emissions. The mean fossil $CO_2$ budget for the remaining 21$^{st}$ century is estimated at 990 Gt for a fifty percent probability of meeting the 2C target [11]. On the RE side, the impact of installations, e.g. in terms of occupied area, should not be environmentally damaging. This criterion appears to be satisfied by the present status of RE technologies, whereas bio-derived fuels, nuclear and hydro do not seem likely to increase much beyond their current levels because of physical and political constraints [12, 13].

Defining the desired level of per capita energy is the key for Guideline II. Energy use per capita is correlated with economic development with high-income countries using on average 6000W compared to a world average of 2250W in 2013. Because fossil fuels have fairly low final energy conversion efficiency it could be assumed that an RE-based, electrified economy will require lower levels of per capita primary energy for the same economic output. This advantage is compensated by the fact that an RE-based energy system would require (i) overcapacity (ii) storage and conversion to high-density carriers, and (iii) additional investment to replace the present social and industrial infrastructure to make it compatible with the characteristics of energy production by renewables but also to support a rising income for the developing world. This leads to a divergence in the estimates of the per capita energy needs ranging from a low 1400W [12] up to 10000W [14, 15]. An average net primary power of 2000W per capita may be considered as a lower limit for maintaining an acceptable quality of life in a technical society [16]. To establish a reference global minimum net-power target curve until 2100, we use the recent UN mid-level population projections [17] since the inertia in population dynamics makes the effect of wealth, and by implication energy availability, quite small in this timeframe.

Guideline III specifies the rate of investment that would provide sufficient net energy (Guideline II) before fossil fuels become unavailable (Guideline I). The amount of energy used in building the RE infrastructure depends on the energy return on energy invested (EROEI)[14] of the RE technology portfolio. While the EROEI of fossil fuels declines as a result of depletion[18], the EROEI of RE tends to increase due to a learning-curve effect although it may be counterbalanced by the eventual saturation of the most favorable sites. The EROEI range reported for RE is large, ranging from less than 5 (e.g. biofuels) to over 50 (hydro) with the technologies most amenable to large-scale expansion, solar and wind from less than 10 to over 40 [19,3] although even lower system values are reported [20].

A measure synthesizing these inputs into the net available energy curve is the renewable energy investment ratio ($\varepsilon$), the fraction of the energy society *chooses* to invest in building renewable energy capital over total available energy. It is possible to build a physical balance model for net power availability that relates ($\varepsilon$) to the system's EROEI, the RE lifetime, and the climate-constrained depletion rate of fossil fuels over time. Solving this constrained equation for a given reference demand curve provides the target value of ($\varepsilon$) and desired installations over time [21].

Figure 1 shows a possible SET trajectory that can provide 2000W net per capita by 2100 based on average RE EROEI values and requiring an increase in the RE investment ratio from less than 0.2% to 5.6% in 2055. Since, the SET trajectory is sensitive to the EROEI, the net power target, and the fossil phase-out path, Figure 2 sketches total RE installed capacity for varying EROEI and final demand under three fossil phase-out strategies that conform to the 990GtCO$_2$ target: a 2023 fossil peak and 2055 phase-out, a 2030 peak and 2045 phase-out, as well as a scenario that allows a partial phase-out by 2040 in order to retain limited fossil energy use until the end of the century.

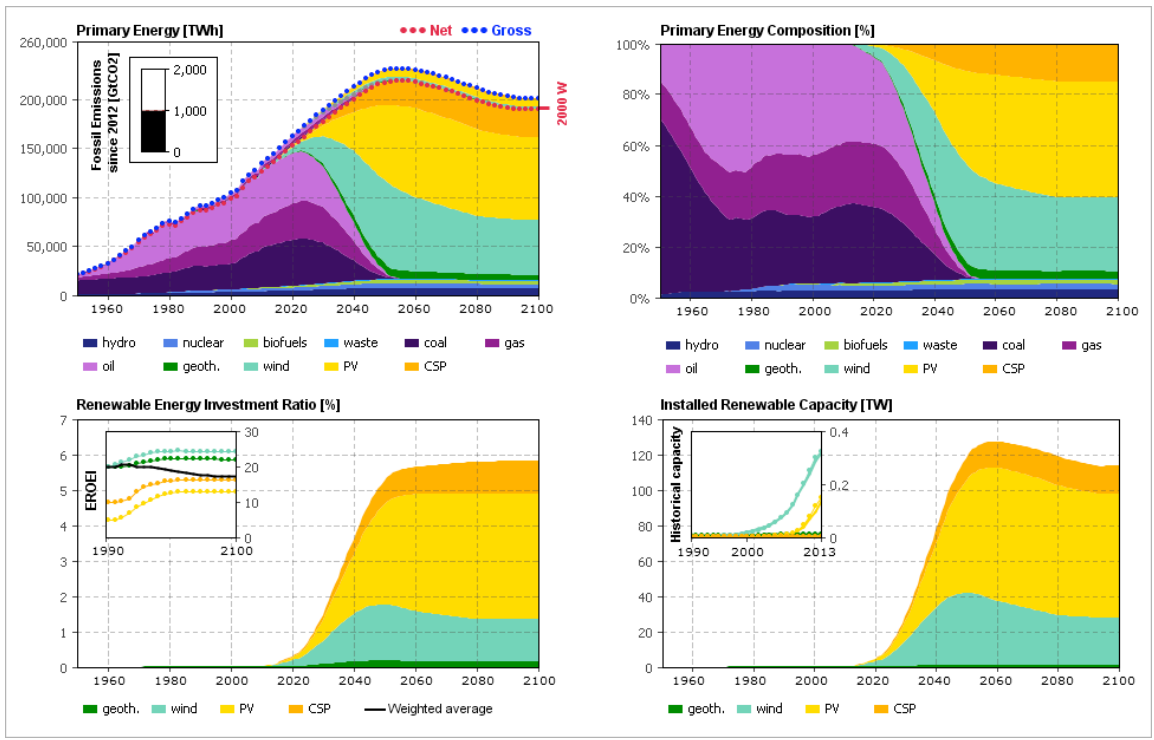

Figure 1: RE portfolio installation trajectory for 2000W net available energy per capita in 2100: (a) Primary Energy. (b) Share of Primary Energy Sources. (c) Renewable Energy Investment Ratio and EROEI Trajectory. (d) Installed RE Capacity (TW).

The slope of Figure 2 contours represents the net capacity addition rate. It is notable that in the critical initial acceleration phase, capacity additions are largely independent of the EROEI and the final demand but critically dependent on the aggressiveness of the required fossil phase-out with corresponding peaks at 5.2, 10.8, and 7.4 TW/year. For reference, in 2013, RE installations were only 0.2TW.

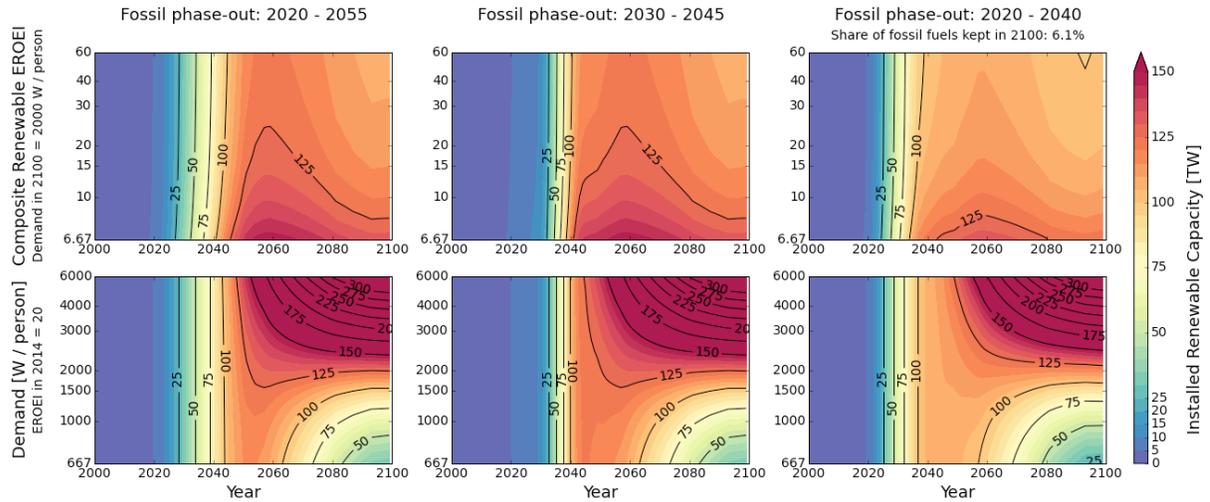

Figure 2: Energy Transition Morphology. Required Installed RE Capacity for Varying EROEI and Net Power Targets under Three Fossil Fuel Phase-out Strategies.

Therefore, a successful SET requires a sustained acceleration in the rate of investment in renewable energy of more than one order of magnitude within the next three decades. A corollary is that a delay in the rise of investments cannot be compensated by subsequent additional acceleration because, if we wait too long, the decline in net energy produced by constrained fossil fuels would become insufficient to power the transition in accordance to the proposed guidelines. In addition, attaining SET rests on replacing a large part of the present capital infrastructure (from appliances to buildings and roads) to utilize the new energy resources. Given that there are significant lead times and many investment decisions taken today (road construction, buildings, aircraft) have useful lives of several decades, it is far from certain that their accelerated depreciation will be acceptable.

These results help set concrete goals for energy planning based on physical reality and its constraints. The question is how (and if) the current economic and political system can generate the mechanisms necessary to reallocate these resources. On this point, we note that the current economy is the result of a reinforcing process in which wealth becomes increasingly abstract and where growth is obtained on the basis of a continuous increase in the primary energy supply. A financial system in which debt is accumulated relying on the belief that future growth will permit to repay it cannot be maintained, given that climate and energy constraints force a steep peak in fossil fuel extraction. One possibility to overcome this problem would be to tie debt extension to the RE investment ratio, as stipulated by Guideline IV, and provide a self-regulating incentive to the financial system to support the energy transition.

The challenge of a sustainable energy transition before he end of the $21^{st}$ century under climate constraints is unprecedented in magnitude, scope, and ambition. It is, nevertheless, doable if we are willing to adopt the "sower's strategy" and invest an appropriate amount of the fossil energy available today into building a sustainable energy future.